# Monopulse-based THz Beam Tracking for Indoor Virtual Reality Applications

Krishan Kumar Tiwari[1] *SMIEEE*, Vladica Sark[1], Eckhard Grass[1, 2], and Rolf Kraemer[1, 3]

[1] IHP - Leibniz-Institut für innovative Mikroelektronik, 15236 Frankfurt (Oder), Germany
[2] Humboldt-Universität zu Berlin, 10099, Berlin, Germany
[3] Brandenburg University of Technology Cottbus-Senftenberg, 03046 Cottbus, Germany
E-mail: {tiwari, sark, grass, kraemer}@ihp-microelectronics.com

## Abstract

Terahertz spectrum is being researched upon to provide ultra-high throughput radio links for indoor applications e.g. virtual reality (VR), etc. as well as outdoor applications e.g. backhaul links, etc. This paper investigates a monopulse-based beam tracking approach for limited mobility users relying on sparse massive multiple input multiple output (MIMO) wireless channels. Owing to the sparsity, beamforming is realized using digitally-controlled radio frequency (RF) / intermediate-frequency (IF) phase shifters with constant amplitude constraint for transmit power compliance. A monopulse-based beam tracking technique, using received signal strength indication (RSSI) is adopted to avoid feedback overheads for obvious reasons of efficacy and resource savings. The Matlab implementation of the beam tracking algorithm is also reported. This Matlab implementation has been kept as general purpose as possible using functions wherein the channel, beamforming codebooks, monopulse comparator, etc. can easily be updated for specific requirements and with minimum code amendments.

## 1 Introduction

Terahertz spectrum (0.1 to 10 THz) provides unallocated large bandwidths needed for ultra-high data rate radio links for indoor applications such as virtual reality (VR) / augmented reality (AR) [1], [2], [3]. Such high throughput wireless communications enable the users to enjoy VR/AR experiences without the need to carry heavy backpacks [4], [5] and tethered connections.

As the wavelengths are very small, large number of antenna elements can be packed in small form-factors, both at indoor access points (APs) and users' headsets. It is indispensable to employ beamforming techniques to realize high gain, narrow beams for communications to be possible despite the large free-space path-loss (FSPL) and raised noised floor due to high carrier frequency and huge bandwidths, respectively. Despite the potential of very high MIMO dimensions, only a few multi-path components (MPCs) actually exist from the transmitter (Tx) to the receiver (Rx) due to electromagnetic propagation conditions at such high frequencies e.g. high atmospheric attenuation, indoor foliage losses, etc. This sparsity of THz MIMO channels is easily evident in beamspace [6], [7], [8].

Such sparse MIMO channels are more easily and more efficiently learnt in beamspace than in the spatial signal space. Exhaustive beam search and hierarchical beam search are the state-of-the-art beam training / channel learning / channel estimation techniques for learning sparse MIMO channels in beamspace [9], [10]. However, beam search processes have significant overheads and losses [11], [12] which can be avoided and instead used for payload data communications, if the need for beam search could be minimized. It is thus imperative to track the user by applying beam tracking algorithms. With effective beam tracking, beam search can be circumvented as long as the track is not lost due to unforeseen / random reasons. There are various ways of tracking users e.g. Kalman filter based tracking, etc. Such techniques need feedback i.e. noisy observations / estimates from the user for tracking. It is desirable not to employ feedback-based tracking for obvious reasons of effectiveness and resource savings.

We opt for Monopulse-based beam tracking as is used for sensing applications such as radars, optical trackers, etc. [13]. This manuscript reports monopulse-based beam tracking for indoor VR applications with sparse massive MIMO channels. A Matlab implementation is also presented and confirms the effectiveness of the proposed approach. The Matlab implementation has been kept as general purpose as possible using functions wherein the channel, beamforming codebooks, monopulse comparator, etc. can easily be updated for specific requirements with minimum code amendments.

The rest of the paper is organized as follows: Section 2 specifies the system model / specifications and tracking requirements, Section 3 provides a brief of monopulse principle, Section 4 presents the design and Matlab implementation of monopulse-based beam tracking, and Section 5 summarizes and concludes the paper.

## 2 Tracking Requirement Specifications

In VR/AR scenarios, users have limited mobility with maximum speed of 4 kilometers per hour (km/h). The typical electronic RF beam switching takes about nanoseconds (ns) which is orders of magnitude faster than the user speed. VR/AR users move inside the VR/AR room on a plane surface. For the sake of simplicity and without the loss of generality, we consider the uniform linear array (ULA) implementation. The angular tracking is needed only in the one dimensional azimuthal axis parallel to the ULA as the beam selections remain the same for user movement along the orthogonal dimension / direction / axis in the azimuthal plane.

Current channel parameters i.e. angle of arrival (AoA), angle of departure (AoD), and channel gain are the inputs to the beam tracking process coming from the results of beam search operation. The MIMO channel can be specified by AoA, AoD, and channel gain [6], [chapters 7, 8 of [14]].

For minimal spatial inter-beam correlation / inter-beam leakage / inter-beam coupling, orthogonal / unitary / Butler matrix / discrete Fourier transform (DFT) beamforming codebooks are used at the transmitter and the receiver [section 3.10 of [15]]. For an N element ULA, N orthogonal beams can be created which divide the beamforming space in N angular sectors corresponding to the half power beam widths (HPBWs) in u-space where u = sin (θ), 'θ' being the angle measured from the ULA broadside.

The beamforming space is 'θ' from $-\pi$ to $+\pi$ radians. As seen in Figure 1, for orthogonal / DFT / unitary beamforming codebooks, at the peak or main response axis (MRA) of one beam all other beams have their nulls i.e. they do not receive or transmit any energy in that direction. The orthogonal beamforming codebook for N=8 element ULA is given in Table 1. For example, the first, fifth, and seventh columns are the beamforming vectors for MRAs at -90, 0, and 48.59 degrees, respectively, measured from the ULA broadside. The magnitudes of the beamforming weights are constant and equal to unity for compliance with the transmit power constraint. However, different beamforming weights have different phase-shifts. These phase-shifts are implemented using digitally-controlled RF/IF phase-shifters to dispense with power-hungry RF chains (RFCs) which would have been needed for the digital / baseband implementation of such beamforming weights. Typically, 3 to 4 control bits are used to set the phase-shift values of the RF/IF phase shifters. DFT codebooks correspond to the left and right unitary matrices obtained after singular value decomposition (SVD) of a MIMO channel matrix. Thus, they are optimum in communication theoretic sense [chapter 7 of [14]].

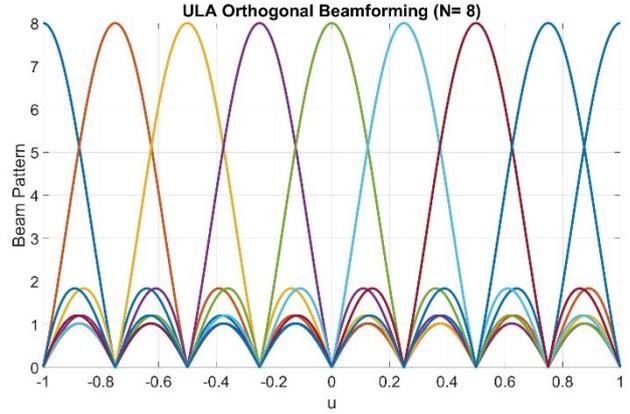

Figure 1. ULA Orthogonal beamforming, N=8

Table 1: 8 pt. ULA DFT Beamforming Codebook

| u=-1 | u=-0.75 | u=-0.5 | u=-0.25 | u=0 | u=0.25 | u=0.5 | u=0.75 |
|---|---|---|---|---|---|---|---|
| 1 | 1 | 1 | 1 | 1 | 1 | 1 | 1 |
| -1 | -0.707+0.707j | j | 0.707+0.707j | 1 | 0.707-0.707j | -j | -0.707-0.707j |
| 1 | -j | -1 | j | 1 | -j | -1 | j |
| -1 | 0.707+0.707j | -j | -0.707+0.707j | 1 | -0.707-0.707j | j | 0.707-0.707j |
| 1 | -1 | 1 | -1 | 1 | -1 | 1 | -1 |
| -1 | 0.707-0.707j | j | -0.707-0.707j | 1 | -0.707+0.707j | -j | 0.707+0.707j |
| 1 | j | -1 | -j | 1 | j | -1 | -j |
| -1 | -0.707-0.707j | -j | 0.707-0.707j | 1 | 0.707+0.707j | j | -0.707+0.707j |

## 3 Monopulse preliminaries

In amplitude comparison monopulse angular tracking [13], RSSI derived from the received electromagnetic (EM) energy / radiation is used to track the object of interest. A multi-lobe antenna is used as the primary sensor to generate the error signal proportional to the angular offset of the target from the boresight of the antenna. As seen in Figure 2, two receive lobes squinted at half power beam width (HPBW) from the antenna boresight are created. The object of interest is assumed to be at the antenna boresight at the beginning of beam tracking. Antenna boresight is denoted as 'B'. If the actual AoA is on the right of 'B' e.g. axis marked as 'R', then the right lobe will capture more received EM energy as compared to the left lobe yielding a positive error signal at the output. On the contrary, for left AoA e.g. 'L', the left lobe will collect more received EM energy than the right lobe giving a negative error signal at the output. This error signal can be used for the angular / beam tracking of a user. The AP receives radio signals from VR / AR users during handshaking / acknowledgment phases of the data communication. Thus, no extra radio reception / feedback is needed for error generation.

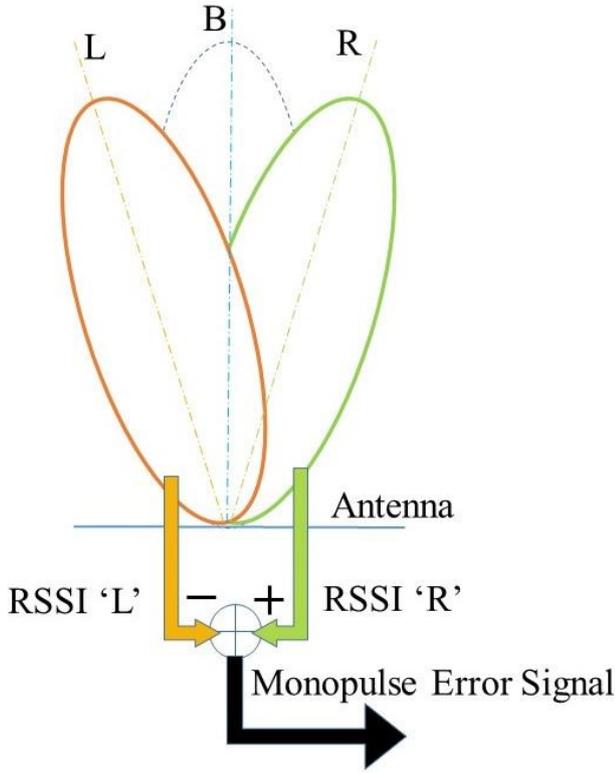

Figure 2. Illustration of monopulse error signal generation

With reference to Figure 2, the error signal magnitude indicates the amount of angular offset. The larger the angular offset, the larger the error signal. The sign of the error i.e. +ve or –ve can be used to determine whether the user is on the right or the left side of the antenna array boresight.

Based on the monopulse error signal, the RF/IF beamforming phase-shift values are set for next possible right or left beam MRA for positive or negative monopulse error signal, respectively. The error is tested iteratively which should go down to almost zero, within computational thresholds, if the new beam MRA matches with the AoA; else the beamforming phase-shift values are updated again to reduce the error.

As seen in equation (1), due to sin of cos of 'θ' and cos of sin of 'θ' terms, the beamwidth increases as the beam is steered away on either side of the ULA boresight. The exponential term on the left hand side is the expression for beamforming weight or phase shift terms in the array manifold vector / steering vector for a ULA [sections 2.2, 2.3 of [15]] assuming narrowband signals. To overcome this issue, we use the HPBWs in $u$-space to derive the value of left and right beam squints because in $u$-space the beamwidth remains constant even with steering [chapters 2, 3 of [15]], where $u = \sin\theta$. As evident in equation (2), in $u$-space, the beamwidth remains constant across the scanning space. $d$ is the ULA inter-element spacing and $\lambda$ is the carrier wavelength.

$$\frac{d}{d\theta} e^{-j\frac{2\pi}{\lambda} d \sin(\theta)} = -\frac{2\pi}{\lambda} d \cos(\theta) \sin\left(\frac{2\pi}{\lambda} d \cos(\theta)\right) - j\frac{2\pi}{\lambda} d \cos(\theta) \cos\left(\frac{2\pi}{\lambda} d \sin(\theta)\right)$$

(Eq. 1)

$$\frac{d}{du} e^{-j\frac{2\pi}{\lambda} d u} = -j\frac{2\pi}{\lambda} d\, e^{-j\frac{2\pi}{\lambda} d u}$$

(Eq. 2)

In case of broadband signals, they can be broken down in the frequency domain into smaller bandwidth sub-channels by using DFT / orthogonal frequency domain multiplexing (OFDM) [sections 3.4.4 and 4.4 of [14]].

Our Matlab codes implement the following:

1. DFT beamforming codebooks are created for Tx & Rx ULAs
2. Beam search o/p is passed as i/p for beam tracking
3. MIMO channel matrix is created based on AoA, AoD, and channel gain
4. Monopulse error is calculated
5. Next left / right beam is selected
6. Steps 4 & 5 are iterated

## 4 Conclusions

THz RF/IF beam tracking requirements specific for indoor VR/AR applications have been identified. Monopulse-based beam tracking has been implemented with discrete beam MRAs. This method provides a minimum overhead solution as it does not need feedback. For systems with more than two RFCs, monopulse-based approach has lower overhead than sequential lobing and without any extra hardware requirement. The error signal reduces progressively to zero in a few iterations and the beam track is maintained consistently. As long as the AoA does not change i.e. the user does not move, the same beam is retained i.e. no steering takes place.

The Matlab codes have used ideal phase-shift values in DFT codebooks. In practice, only discrete phase-shift values will be available with digitally-controlled phase-shifters. An impact analysis and validation of quantized phase values is an interesting future work with practical relevance.

# 5 Acknowledgment

This work has received funding from the European Union's Horizon 2020 research and innovation programme under grant agreement No 761329 (WORTECS).

# 6 References

[1] R. Piesiewicz et al., "Short-Range Ultra-Broadband Terahertz Communications: Concepts and Perspectives," *IEEE Antennas and Propagation Magazine*, vol. 49, no. 6, pp. 24-39, Dec. 2007.

[2] K. Huang and Z. Wang, "Terahertz Terabit Wireless Communication," *IEEE Microwave Magazine*, vol. 12, no. 4, pp. 108-116, June 2011.

[3] I. F. Akyildiz, J. M. Jornet, and C. Han, "Terahertz Band: Next Frontier for Wireless Communications," *Phys. Commun.*, vol. 12, no. 2, pp. 16-32, Sept. 2014.

[4] EU WORTECS Consortium, "European Union's Horizon 2020 research and innovation programme under grant agreement No 761329 WORTECS project website," [Online]. Available: https://wortecs.eurestools.eu/ [Accessed: 21-Mar-2019].

[5] M. Badawi et. al, "EU H2020 WORTECS Deliverable D2.3: Focus on Virtual Reality," [Online]. Available: https://wortecs.eurestools.eu/deliverables-dissemination/ [Accessed: 21-Mar-2019].

[6] A. M. Sayeed, "Deconstructing multiantenna fading channels," *IEEE Transactions on Signal Processing*, vol. 50, no. 10, pp. 2563-2579, Oct. 2002.

[7] A. Sayeed and J. Brady, "Beamspace MIMO for high-dimensional multiuser communication at millimeter-wave frequencies," *IEEE Global Communications Conference (GLOBECOM)*, Atlanta, GA, pages 3679-3684, Dec. 2013.

[8] G. H. Song, J. Brady, and A. Sayeed, "Beamspace MIMO transceivers for low-complexity and near-optimal communication at mm-wave frequencies," IEEE International Conference on Acoustics, Speech and Signal Processing (ICASSP), Vancouver, BC, pages 4394-4398, May 2013.

[9] S. Hur, T. Kim, D. J. Love, J. V. Krogmeier, T. A. Thomas, and A. Ghosh, "Millimeter Wave Beamforming for Wireless Backhaul and Access in Small Cell Networks," IEEE Transactions on Communications, vol. 61, no. 10, pp. 4391-4403, Oct. 2013.

[10] S. Hur, T. Kim, D. J. Love, J. V. Krogmeier, T. A. Thomas, and A. Ghosh, "Multilevel millimeter wave beamforming for wireless backhaul," IEEE GLOBECOM Workshops (GC Wkshps), Houston, Texas, USA, pages 253-257, Dec. 2011.

[11] K. K. Tiwari, J. S. Thompson, and E. Grass, "Noise performance of Orthogonal RF beamforming for millimetre wave massive MIMO communication systems," The Tenth International Conference of Wireless Communications and Signal Processing (WCSP), Hangzhou, China, Oct. 2018, pp. 1-7.

[12] K. K. Tiwari, E. Grass, and R. Kraemer, "Noise performance of Orthogonal RF beamforming for THz Radio Communications," The Ninth IEEE Annual Computing and Communication Workshop and Conference (CCWC), University of Nevada, Las Vegas, USA, Jan. 2019, pp. 746-751.

[13] Samuel M. Sherman and David K. Barton, "Monopulse Principles and Techniques," *Artech Publishing House*, 2011.

[14] D. Tse and P. Viswanath, "Fundamentals of Wireless Communication," *Cambridge University Press*, 2005.

[15] H. L. V. Trees, Optimum Array Processing: Part IV of Detection, Estimation, and Modulation Theory, *New York: Wiley-Interscience*, 2002.